\documentclass[conference]{IEEEtran}
\ifCLASSINFOpdf
\else
\fi
\usepackage{amsmath,dsfont,bbm,epsfig,amssymb,amsfonts,amstext,verbatim,amsopn,cite}
\usepackage{subfigure,multirow,multicol,lipsum,xfrac}
\usepackage{amsthm}
\usepackage{mathtools,amsthm}
\usepackage{perpage}
\usepackage{balance}
\usepackage{url}
\usepackage{amsfonts}
\usepackage{epsfig}
\usepackage[font={small}]{caption}
\usepackage{psfrag}	
\usepackage{etoolbox}
\usepackage{algorithmicx}
\usepackage[Algorithm,ruled]{algorithm}
\usepackage{algpseudocode}
\usepackage{pifont}
\usepackage[utf8]{inputenc}
\usepackage[T1]{fontenc}  
\usepackage[nolist]{acronym}
\MakePerPage{footnote}
\usepackage{textcomp}
\usepackage{paralist}
\usepackage{enumitem}
\usepackage{bbm}
\usepackage[process=auto]{pstool}
\usepackage{tikz,pgfplots}
\usetikzlibrary{shapes,arrows}
\usetikzlibrary{fit}
\tikzset{%
  highlight/.style={rectangle,rounded corners,fill=red!15,draw,fill opacity=0.5,thick,inner sep=0pt}
}

\newcommand{\setM}{\mathbb{M}}
\newcommand{\setA}{\mathbb{A}}

\newcommand{\setR}{\mathbb{R}}

\newcommand{\setS}{\mathbb{S}}
\newcommand{\setW}{\mathbb{W}}

\newcommand{\setO}{\mathbb{O}}
\newcommand{\setF}{\mathbb{F}}

\newcommand{\setI}{\mathbb{I}}

\DeclareMathOperator*{\argmin}{argmin}

\newcommand{\man}{\mathcal{N}}

\newcommand{\bd}{{\mathbf{d}}}

\newcommand{\bx}{{\boldsymbol{x}}}

\newcommand{\set}[1]{\left\lbrace#1\right\rbrace}
\newcommand{\brc}[1]{\left( #1 \right)}
\newcommand{\dbc}[1]{\left[ #1 \right]}

\newcommand{\bz}{{\boldsymbol{z}}}

\newcommand{\bs}{{\boldsymbol{s}}}

\newcommand{\bv}{{\boldsymbol{v}}}

\newcommand{\bw}{{\boldsymbol{w}}}

\newcommand{\bgg}{{\mathbf{g}}}

\newcommand{\dif}{\mathrm{d}}

\newcommand{\by}{{\boldsymbol{y}}}

\newcommand{\trp}{\mathsf{T}}

\newcommand{\mA}{\mathbf{A}}

\newcommand{\mI}{\mathbf{I}}
\newcommand{\mone}{\mathbf{1}}

\newcommand{\mU}{\mathbf{U}}

\newcommand{\mY}{\mathbf{Y}}

\newcommand{\E}{\mathbb{E}\hspace{.5mm}}

\newcommand{\norm}[1]{\lVert #1 \rVert}

\newcommand{\abs}[1]{\lvert #1 \rvert}

\newtheoremstyle{mystyle}%                % Name
  {}%                                     % Space above
  {}%                                     % Space below
  {}%                                     % Body font
  {}%                                     % Indent amount
  {\bfseries}%                            % Theorem head font
  {:}%                                    % Punctuation after theorem head
  { }%                                    % Space after theorem head, ' ', or \newline
  {}%                                     % Theorem head spec (can be left empty, meaning `normal')
\theoremstyle{mystyle}

\newtheorem{definition}{Definition}

\algnewcommand\algorithmicLet{\textbf{Let}}
\algnewcommand\Let{\item[\algorithmicLet]}
\algnewcommand\algorithmicSet{\textbf{Set}}
\algnewcommand\Set{\item[\algorithmicSet]}

\algnewcommand\algorithmicInitiate{\textbf{Initiate}}
\algnewcommand\Initiate{\item[\algorithmicInitiate]}
\algnewcommand\algorithmicStart{\textbf{Begin}}
\algnewcommand\Begin{\item[\algorithmicStart]}
\algnewcommand\algorithmicEnd{\textbf{End}}
\algnewcommand\End{\item[\algorithmicEnd]}

\algnewcommand\algorithmicOutP{\textbf{Output:}}
\algnewcommand\Out{\item[\algorithmicOutP]}

\algnewcommand\algorithmicInP{\textbf{Input:}}
\algnewcommand\In{\item[\algorithmicInP]}

\newcounter{bar}

\begin{document}
\title{An Adaptive Bayesian Framework for Recovery of Sources with Structured Sparsity}

% Authors
\author{
\IEEEauthorblockN{
Ali Bereyhi and
Ralf R. M\"uller
}
\IEEEauthorblockA{
%Institute for Digital Communications, 
Friedrich-Alexander Universit\"at Erlangen-N\"urnberg, Germany\\
\{ali.bereyhi, ralf.r.mueller\}@fau.de
\thanks{This work has been accepted for presentation in 2019 IEEE International Workshop on Computational Advances in Multi-Sensor Adaptive Processing (CAMSAP). The link to the final version in the proceedings will be available later.}
}
}


\IEEEoverridecommandlockouts

\maketitle

\begin{acronym}
\acro{oas}[OAS]{oversampled adaptive sensing}
\acro{awgn}[AWGN]{additive white Gaussian noise}
\acro{iid}[i.i.d.]{independent and identically~dis-tributed}
\acro{rhs}[r.h.s.]{right hand side}
\acro{lhs}[l.h.s.]{left hand side}
\acro{wrt}[w.r.t.]{with respect to}
\acro{mse}[MSE]{mean squared error}
\acro{mmse}[MMSE]{minimum MSE}
\acro{snr}[SNR]{signal-to-noise ratio}
\acro{sinr}[SINR]{signal to interference and noise ratio}
\acro{mf}[MF]{match filtering}
\end{acronym}

\begin{abstract}
In oversampled adaptive sensing (OAS), noisy measurements are collected in multiple subframes. The sensing basis in each subframe is adapted according to some posterior information exploited from previous measurements. The framework is shown to significantly outperform the classic non-adaptive~com-pressive sensing approach.

This paper extends the notion of OAS to signals with structur-ed sparsity. We develop a low-complexity OAS algorithm based on structured orthogonal sensing. Our investigations depict that the proposed algorithm outperforms the conventional non-adaptive compressive sensing framework with group LASSO recovery via a rather small number of subframes.

%is a recently proposed Bayesian framework which sequentially adapts the sensing basis. In OAS, estimation quality is, in each step,~measured by conditional mean squared errors (MSEs), and~the~basis for the next sensing step is adapted accordingly.~For~given\\ average sensing time, OAS reduces the MSE compared~to~non- adaptive schemes, when the signal is sparse. This~paper~studies the asymptotic performance of Bayesian OAS,~for~unitarily~in- variant random projections. For sparse signals, it is shown that OAS with Bayesian recovery~and~hard~adaptation significantly outperforms the minimum MSE bound for non-adaptive sensing. To address implementational aspects, two computationally tractable algorithms are proposed, and their performances are compared against the state-of-the-art non-adaptive algorithms via numerical simulations. Investigations depict that these low-complexity OAS algorithms, despite their suboptimality,~outperform well-known non-adaptive schemes for sparse recovery, such as LASSO, with rather small oversampling factors. This gain grows, as the compression rate increases.\vspace*{-2mm}
\end{abstract}


\IEEEpeerreviewmaketitle

\section{Introduction}
\label{sec:intro}
The recently proposed \ac{oas} framework has shown privileged performance for \textit{time-limited} sensing in noisy environments \cite{muller2018oversampled,muller2018randomoversampled}. Unlike earlier adaptive approaches, e.g., \cite{haupt2009compressive,haupt2009adaptive,malloy2014near}, this scheme allows for \textit{oversampling}. In this scheme,~the~signal is sensed in multiple steps, referred to as \textit{subframes}. The sensing matrix in each~subframe~is~ada-pted based on some posterior information determined from the measurements of previous subframes. In \cite{muller2018oversampled}, it has been demonstrated that \ac{oas} achieves a considerable performance gain, when some prior information on the signal is available. The most well-known form of such prior information~is~\textit{sparsity} which was explicitly studied in \cite{muller2018oversampled,muller2018randomoversampled}. Investigations have depicted that even \textit{suboptimal low-complexity} \ac{oas} algorithms outperform well-known non-adaptive~compressive~sensing techniques in time-limited scenarios. This is~intuitively~il-lustrated as follows: When the signal is sparse, zero samples are detected in initial subframes even by low-quality measurements. These samples are then excluded in~next~subframes, where we focus on sensing the non-zero samples.

The previous studies on the \ac{oas} framework model~the~samples of a sparse signal as an \ac{iid} process which does not consider any structure on the sparsity. It is however known that in many applications with sparse signals, the samples have structural dependencies, e.g., \cite{he2009exploiting,baraniuk2008model}. In such applications, the recovery performance of conventional compressive sensing techniques can be improved by taking into account the sparsity structure \cite{baraniuk2008model,eldar2009robust,duarte2011structured,eldar2010block}.

From Bayesian points of view, structured sparsity provides further prior information on the signal. This intuitively implies that \ac{oas} achieves higher performance gains when it is employed to sense signals with structured sparsity. In this work, we aim to study the performance of \ac{oas} in such scenarios. To this end, we develop a low-complexity \ac{oas} scheme based on structured orthogonal sensing. Our investigations show that the proposed adaptive scheme with few subframes significantly outperforms the non-adaptive state-of-the-art.

\textit{Notation:} Scalars, vectors and matrices are shown with non-bold, bold lower case and bold upper case letters, respectively. $\mI_K$ and $\boldsymbol{0}_{K\times N}$ are the $K \times K$ identity matrix and $K \times N$~all-zero matrix, respectively. $\mA^{\trp}$ denotes the transpose of $\mA$. The set of real numbers is shown by $\setR$. We use the shortened~notation $[N]$ to represent $\set{1, \ldots , N}$.

\section{Problem Formulation}
We consider a sensing setup in which a vector of $N$ signal samples collected in $\bx\in\setR^{N}$ is to be sensed via $K$ distinct sensors within a fixed time interval of duration $T$. The sensing process is assumed to be linear and noisy. Hence, the vector of measurements collected by the sensor network within $t\leq T$ seconds of sensing is represented as $\by = \left. \mA \right. \bx + \bz$, where $\mA \in \setR^{K\times N}$ is the sensing matrix whose entries~are~tunable, and $\bz \in \setR^{K}$ denotes additive white Gaussian noise with zero mean and variance $\sigma^2\brc{t}$. The dependency of the noise variance on the sensing time models the sensing quality.

\subsection{Model for Time-Limited Sensing} 
As indicated, the sensing process is to be limited to a time duration of $T$. It can hence be performed either in one step for a duration of $T$ or in $M$ steps each lasting~for~$T/M$. In the former case, the sensing process ends with $K$ noisy measurements; however, the latter scheme collects $MK$ measurements in total. Intuitively, the quality of measurements obtained by the first approach is higher than those acquired via multiple sensing steps. We model this phenomenon by setting the noise variance reversely proportional to the sensing time, i.e., for sensing duration $t$, $\sigma^2\brc{t} = {\sigma^2_0}/ {t}$ with $\sigma^2_0$ denoting the variance of noise within a unit of time. 

This model is straightforwardly justified for various types of sensing devices following the corresponding circuitry models; see \cite{muller2018oversampled} for some detailed discussions. From systematic viewpoint, this model agrees with the physical intuition, since the \ac{snr} at each sensor grows linearly with time. Considering this model, there is a trade-off between the number of total collected measurements and the sensing quality. More measurements are acquired at the expense of shorter sensing time which results in higher noise variance.

\subsection{Bayesian OAS Framework}
The Bayesian \ac{oas} framework, introduced and analyzed in \cite{muller2018oversampled,muller2018randomoversampled}, refers to the following sequential sensing procedure:
\begin{enumerate}[label=(\alph*)]
\item The sensing time $T$ is divided to $M$ \textit{subframes}.
\item In subframe $m\in\dbc{M}$, the sensors measure
\begin{align}
\by_m = \left. \mA_m \right. \bx + \bz_m \label{eq:y_m}
\end{align}
for some sensing matrix $\mA_m$ and measuring noise $\bz_m\sim \mathcal{N} \brc{ \boldsymbol{0}, \sigma^2_{\rm sub} \mI_K }$, where $\sigma^2_{\rm sub} = \sigma^2\brc{T/M} = M\sigma^2\brc{T}$.
\item From the stacked measurements in subframe $m$, i.e.,
\begin{align}
\mY_m \coloneqq \dbc{ \by_1, \ldots, \by_m},
\end{align}
a \textit{Bayesian estimation} of the samples is determined as
\begin{align}
\hat{\bx}_m  = \E \set{\bx | \mY_m , \setA_m},
\end{align}
where %$\setA_m$ is %denotes the collection of sensing matrices in subframe $m$, i.e.,
%\begin{align}
$\setA_m = \set{ \mA_1, \ldots,\mA_m }$,
%\end{align}
and the expectation is taken with respect to some \textit{postulated} prior distribution $q\brc{\bx}$.
\item Given the estimation in subframe $m$, the vector of \textit{posterior information} is determined as
\begin{align}
\bd_m = \E \set{\dif \dbc{\bx; \hat{\bx}_m} | \mY_m , \setA_m},
\end{align}
for some distortion function $\dif \dbc{\cdot; \cdot}$. In general, the dimension of the posterior information vector can be different from the signal dimension. We hence denote it by $B$, i.e., $\bd_m \in \setR^{B}$, to keep the formulation generic.
\item The sensor network constructs the sensing matrix of the next subframe based on $\bd_m$, i.e., $\mA_{m+1} = f_{\rm Adp} \brc{\bd_m}$~for some adaptation function $f_{\rm Adp} \brc\cdot$.
\end{enumerate}

%In \cite{muller2018oversampled,muller2018randomoversampled}, it has been demonstrated that \ac{oas} achieves a considerable performance gain, when some prior information on the signal is available\footnote{This is modeled via the postulated prior $q\brc{\bx}$ in the \ac{oas} formulation.}. The most well-known form of prior information in the literature is \textit{sparsity}. The initial study has depicted that the \ac{oas} framework outperforms the~conventio-nal compressive sensing techniques in time limited scenarios. This result is intuitively illustrated as follows: When the signal is sparse, the zero samples are detected in initial subframes even by low-quality measurements. These samples are then excluded in the remaining subframes. For non-zero samples, the impact of higher noise variance  in each subframe is further compensated via an effective combining approach.

\subsection{Signals with Structured Sparsity}
\label{sec:signal}
We assume that the signal samples have a structured sparsity pattern. To model the signal, we follow the generic \textit{structured sparsity} model introduced in \cite[Definition 2]{baraniuk2008model}: For $L\leq N$, let $\setI \subseteq \dbc{N}$ be a subset of $L$ indices, i.e., $\abs{\setI} = L$. Define $\bx_\setI \in \setR^L$ to be a vector constructed by collecting those entries in $\bx$ whose indices are in $\setI$. Then, $\setS_\setI$ is said to be a \textit{canonical $L$-sparse subspace} corresponding to index subset $\setI$, when
\begin{align}
\setS_\setI = \set{ \bx: \bx_\setI \in \setR^L \text{ and } \bx_{\setI^C} = \boldsymbol{0}_{N-L} }.
\end{align}

Assume $\setS$ is a subspace which is partitioned into $S$ canonical $L$-sparse subspaces, i.e., %
%\begin{align}
$\setS = \cup_{s=1}^S \setS_{\setI_s}$
%\end{align}
for some distinct index subsets $\setI_1, \ldots, \setI_S$. In this case, $\setS$ is said to represent a \textit{structured sparsity model} with sparsity $L$ on a union of $S$ sparse subspaces. Examples of structured sparsity models are \textit{tree-based} and \textit{block sparse} signals \cite{baraniuk2008model,duarte2011structured,yuan2006model}.

An stochastic model for structured sparsity can be described by a prior distribution for which we have $\Pr\set{\bx\notin \setS} = 0$. In the sequel, we give a stochastic model for the specific example of block sparsity. We use this model later to investigate~our~approach. For sake of simplicity, we present the model for sparse signals whose blocks are of similar size. Extensions to signals consisting of blocks with various lengths is straightforward.

\begin{definition}[Random block sparse model] 
Let $L$ be a divisor of $N$ and define $B=N/L$. $\bx$ is said to be block sparse with block length $L$ and sparsity factor $\xi$, when for $b\in \dbc{B}$
\begin{align*}
\bx_b = \dbc{x_{\brc{b-1}L+1}, \ldots, x_{bL}}^\trp
\end{align*}
reads $\bx_b = \psi_b \bs_b $ with $\bs_b\in\setR^L$ being a~\textit{continuous}~random vector and $\psi_b$ being a $\xi$-Bernoulli random variable, i.e.
\begin{align*}
\Pr\set{\psi_b = 1} = 1- \Pr\set{\psi_b = 0} = \xi.
\end{align*}
\end{definition}

The above model consists of $B$ blocks of length $L$, each of them being either a vector of all zeros or completely non-zero. Hence, knowing only one sample in each block, one can recover the \textit{support} of $\bx$. For large $N$, the fraction of non-zero blocks is $\xi$ which equals the fraction non-zero samples.

\subsection{Objectives and Performance Measure}
The main objective of this study is to investigate the impact of sparsity structure on the performance of \ac{oas}. To this end, we consider the following metric to quantify the performance:

\begin{definition}[Average distortion]
Let $\mY \in \setR^{K\times M}$, for some integer $M$, contain all measurements of $\bx$ collected within~the restricted sensing time $T$. Assume $\bgg\brc\cdot : \setR^{K\times M} \mapsto \setR^N$ denotes an algorithm~recovering the samples in $\bx$ as $\hat{\bx} = \bgg\brc\mY$. The average distortion $D$ with respect to the distortion function $\Delta\brc{\cdot;\cdot} : \setR^N \times \setR^N \mapsto \setR^+$ is  defined as %
%\begin{align}
$D = \E \set{\Delta\brc{\bx; \hat{\bx}}}$.
%\end{align}
\end{definition}

\section{Block-Wise OAS via Orthogonal Sensing}
The complexity of the \ac{oas} framework mainly depends on two factors: the \textit{ensemble} from which the sensing matrix is chosen, and the \textit{postulated prior} which is used for estimation in each subframe. On one hand, one can set the postulated prior distribution to the true one and search in each subframe for the optimal sensing matrix for the next subframe.~This~approach results in optimal performance which is achieved at the expense of high computational complexity. On the other hand, one may restrict the ensemble of sensing matrices and/or postulate a different prior distribution, such that the estimation and sensing matrix construction is addressed in each subframe with low complexity. The investigations in \cite{muller2018oversampled} and \cite{muller2018randomoversampled} show that even by following the latter suboptimal approach, the \ac{oas} framework outperforms the benchmark. 

In the sequel, we develop a low complexity \ac{oas} algorithm for recovery of signals with structured sparsity. The algorithm selects the sensing matrix of each subframe from a certain class of \textit{row-orthogonal} matrices. This restriction significantly simplifies the Bayesian estimation in each subframe. For~sake of brevity, we restrict the derivations to signals with block sparsity whose non-zero samples are \ac{iid} Gaussian:~We~assume that $\bx$ is a random block sparse vector with $B=N/L$ blocks of length $L$ in which $\bs_b\sim \mathcal{N}\brc{\boldsymbol{0}, \mI_L}$ for $b\in\dbc{B}$. The framework is however extendable to other stochastic structured sparsity models with straightforward modifications. 

\subsection{Block-wise Orthogonal Sensing Matrices}
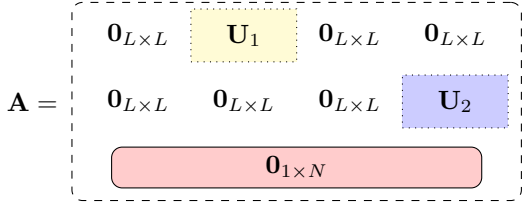
\begin{figure}[t]
\tikzstyle{block} = [draw,rectangle, minimum height=5.5em, rounded corners, minimum width=5em]
\tikzstyle{blockdata} = [draw,rectangle, minimum height=5.5em, minimum width=5em]
\tikzstyle{margin} = [draw, dotted, rectangle, minimum height=2.1em, minimum width=2em]
\tikzstyle{sum} = [draw, circle, minimum size=.3em, node distance=1cm, inner sep=0pt]
\tikzstyle{input} = [coordinate]
\tikzstyle{output} = [coordinate]
\tikzstyle{pinstyle} = [pin edge={to-,thick,black}]
\tikzstyle{iden} = [draw, rectangle, dashed, rounded corners]
\begin{center}
\begin{tikzpicture}[auto,node distance=2.5cm,>=latex']
    \node [input, name=start] { };
    \node [input, above of=start, name=line1,node distance=1.25em] { };
    \node [input, below of=start, name=line2,node distance=1.25em] { };
    \node [input, below of=start, name=line3,node distance=3.75em] { };

\node [iden, right of=line2, minimum width=17em,  minimum height=7.5em, node distance=7em] (Mat) { };    
    
    \node [pinstyle, right of=line1, node distance=1em] (O1) {$\boldsymbol{0}_{L\times L}$};
    \node [margin, right of=line1, minimum width=4em,  minimum height=2em, node distance=5em, fill=yellow!20] (U_1) {$\mU_1$};
    %\node [pinstyle, right of=line1, node distance=5em] (O2) {$\mU_1$};
    \node [pinstyle, right of=line1, node distance=9em] (O3) {$\boldsymbol{0}_{L\times L}$};
    \node [pinstyle, right of=line1, node distance=13em] (O4) {$\boldsymbol{0}_{L\times L}$};
    
    \node [pinstyle, right of=line2, node distance=1em] (U1) {$\boldsymbol{0}_{L\times L}$};
    \node [pinstyle, right of=line2, node distance=5em] (U2) {$\boldsymbol{0}_{L\times L}$};
    \node [pinstyle, right of=line2, node distance=9em] (U3) {$\boldsymbol{0}_{L\times L}$};
    \node [margin, right of=line2, minimum width=4em,  minimum height=2em, node distance=13em, fill=blue!20] (U_2) {$\mU_2$};
    %\node [pinstyle, right of=line2, node distance=13em] (U4) {$\mU_2$};
    
    \node [block, right of=line3, minimum width=14em,  minimum height=1em, node distance=7em, fill=red!20] (last row) {$\boldsymbol{0}_{1\times N}$};
    
   % \node [pinstyle, right of=line3, node distance=7em] (T) {$\boldsymbol{0}_{1\times N}$};
    
    \node [pinstyle, left of=line2, node distance=3em] (T) {$\mA=$};

	\end{tikzpicture}
\end{center}
\caption{An example of block-wise orthogonal matrices.\vspace*{-5mm}}
\label{fig:mat}
\end{figure}

We start the derivations by defining a simple class of \textit{block-wise orthogonal} matrices. This class comprises $F \coloneqq \lfloor K/L \rfloor$ orthogonal \textit{principles}; namely, $\mU_f \in \setR^{L\times L}$ for $f\in \dbc{F}$ which satisfy $\mU_f \mU_f^\trp = \mI_L$. Let %
%\begin{align}
$\setF = \set{i_1 , \ldots , i_F} \subset \dbc{B}$
%\end{align}
be a subset with $F$ distinct indices. The block-wise orthogonal matrix $\mA \in \setR^{K \times N}$ corresponding to $\setF$ is then constructed by setting %the entries of $\mA$ to
\begin{align}
\mA \brc{ \brc{f-1}L + \ell , \brc{i_f-1}L + \lambda} = \mU_f \brc{\ell , \lambda}
\end{align}
for $f\in \dbc{F}$ and $\ell,\lambda \in \dbc{L}$, and the other entries zero.~An~example of block-wise orthogonal matrices with $B=4$ blocks of length $L$ and $K=2L+1$ rows is given in Fig.~\ref{fig:mat}. In this example, $F=2$ and $\setF = \set{2,4}$. Hence, the entries in the last row are zero. $\mU_1,\mU_2 \in \setR^{L\times L}$ denote the principles.

We define $\setO_F$ to be the set of all possible~block-wise~matrices constructed by the principles $\mU_f$ for $f\in \dbc{F}$~from~$F$ distinct indices. %By definition, it is concluded that 
%\begin{align*}
%\abs{\setO_F} = {B \choose F}.
%\end{align*}
The main property of $\mA\in\setO_F$ is that the entries of $\mA \bx$ are partitioned into $F$ blocks in which each block reads $\mU_f \bx_b$~for some $f\in \dbc{F}$ and $b\in\dbc{B}$. It is further straightforward~to~show that for $\mA\in\setO_F$ corresponding to $\setF$ 
\begin{align}
\bv = \mA^\trp \mA \bx = \dbc{ \bv_1,\ldots,\bv_B }^\trp \label{eq:property}
\end{align}
where $\bv_b = \bx_b \mone\set{b\in\setF}$.

By restricting the sensing matrices to be chosen from $\setO_F$, Bayesian estimation and derivation of the posterior information become computationally tractable tasks. In the sequel, we derive  these parameters for the given block sparse model.

\subsection{Bayesian Estimator}
Consider the Bayesian OAS framework, and let $\mA_m\in \setO_F$ be the sensing matrix in subframe $m\in \dbc{M}$. The vector of measurements in this subframe is therefore given by \eqref{eq:y_m}. With straightforward lines of derivations, it is shown that
\begin{align}
\bw_m = \mA_m^\trp \by_m  &= \mA_m^\trp\mA_m \bx + \mA_m^\trp \bz_m%\\
%&= \mA_m^\trp\mA_m \bx + \tilde\bz_m
\end{align}
is a sufficient statistic. Hence, %we have
\begin{align}
\hat{\bx}_m  = \E \set{\bx | \mY_m , \setA_m} = \E \set{\bx | \bw_1 , \ldots, \bw_m}.
\end{align}
Since the blocks are independent, we have%From the independency of blocks in the stochastic block sparsity model, one can conclude that
\begin{align}
\hat{\bx}_{b, m}  = \E \set{\bx_b | \bw_{b,1} , \ldots, \bw_{b,m}}
\end{align}
where $\hat{\bx}_{b, m},\bw_{b,m} \in \setR^L$ denote the $b$-th block of $\hat{\bx}_m$ and $\bw_m$ for $b\in\dbc{B}$, respectively. To continue with derivations, let us define the following two notations:
\begin{itemize}
\item $\setF_m \subseteq \dbc{B}$ represents the index set corresponding to~$\mA_m$. This set contains indices of the blocks whose samples are sensed in subframe $m$.
\item $\setM_b\brc{m} \subseteq \dbc{m}$ contains indices of all subframes at which block $b$ is sensed, i.e. %
%\begin{align}
$\setM_b\brc{m} = \set{ i \in \dbc{m} : b \in \setF_i }$.
%\end{align}
\end{itemize}
By these definition, the Bayesian estimator further reads
\begin{align}
\hat{\bx}_{b, m}  = \E \set{\bx_b | \setW_{b}\brc{m}}
\end{align}
where $\setW_b \brc{m} = \set{ \bw_{b,i}: i \in \setM_b\brc{m} } $. 

Following the property of $\setO_F$ given in \eqref{eq:property}, it is concluded that for $i \in \setM_b\brc{m}$, we have $\bw_{b,i} = \bx_b + \tilde\bz_{b,i}$, where $\tilde\bz_{b,i}\in\setR^{L\times L}$ denotes the $b$-th block of $\mA_i^\trp \bz_i$. This concludes
\begin{align}
\bar{\bw}_b\brc{m} = \sum_{i\in\setM_b\brc{m}} \bw_{b,i} = \abs{\setM_b\brc{m}} \bx_{b} + \sum_{i\in\setM_b\brc{m}} \tilde\bz_{b,i}
\end{align}
is a sufficient static for estimating $\bx_b$. Considering the structure of $\mA_i$, one can conclude that for $i \in \setM_b\brc{m}$, the noise term reads $\tilde\bz_{b,i} = \mU_f^\trp \bz_i^0$ for some principle $\mU_f$ and some $\bz_i^0 \sim \man\brc{\boldsymbol{0}, M\sigma^2\brc{T} \mI_L}$. Hence, we can write
\begin{align}
\bar{\bw}_b\brc{m} = \abs{\setM_b\brc{m}} \bx_{b} + \bar\bz_{b}
\end{align}
where $\bar\bz_b \sim \man\brc{\boldsymbol{0}, \sigma_{b,m}^2 \mI_L}$ with $\sigma_{b,m}^2 \coloneqq \abs{\setM_b\brc{m}} M\sigma^2\brc{T}$.

By substituting the true prior of the block sparse signal, the Bayesian estimator in subframe $m$ reduces to
\begin{align}
\hat{\bx}_{b, m}  = \E \set{\bx_b | \bar{\bw}_b\brc{m} } = \abs{\setM_b\brc{m}} \frac{ \bar{\bw}_b\brc{m}}{C\brc{\bar{\bw}_b\brc{m}}} \label{eq:hat_x_b}
\end{align}
where function $C\brc\cdot: \setR^L \mapsto \setR^+$ reads
\begin{align}
C\brc{\by} \coloneqq V_b\brc{m} \brc{1 +  \frac{ \brc{{1-\xi}} \phi\brc{\by \vert \sigma_{b,m}^2 } }{ \xi \phi\brc{\by \vert V_b\brc{m} }}} \label{eq:C}
\end{align}
with $V_b\brc{m} \coloneqq {\abs{\setM_b\brc{m}}^2 + \sigma_{b,m}^2}$, and $\phi\brc{\by \vert \sigma^2}$ denoting the distribution of a zero-mean Gaussian random vector~with~covariance matrix $\sigma^2 \mI_L$.

\subsection{Posterior Information and Adaptation}
The sensing matrix of each subframe is adapted via~an~adaption function based on the posterior information obtained~in~the previous subframe. A common choice for the posterior information in Bayesian \ac{oas} is the \textit{posterior \ac{mse}} which in the most basic case is determined for each sample of the signal. In order to exploit the sparsity structure of block sparse signals, we set the posterior information to be a $B$-dimensional vector, i.e., $\bd_m = \dbc{d_{1,m}, \ldots, d_{B,m}}^\trp$, whose $b$-th entry is the posterior \ac{mse} of block $b$ in subframe $m$, i.e.,
\begin{align}
d_{b, m}  &= \E \set{\norm{\bx_b- \hat{\bx}_{b, m}}^2 | \bar{\bw}_b\brc{m} } 
\end{align}
By substituting \eqref{eq:hat_x_b} into the definition, we have
\begin{align}
d_{b, m}  = \frac{1}{C\brc{\bar{\bw}_b\brc{m} } } \brc{ \sigma_{b,m}^2 - \frac{ \abs{\setM_b\brc{m}}^2 \norm{\bar{\bw}_b\brc{m}}^2}{C\brc{\bar{\bw}_b\brc{m}}} }. \label{eq:d_m}
\end{align}

The posterior information $\bd_m$ is given to~an~adaption~function which constructs the sensing matrix in the next subframe, i.e., $\mA_{m+1}$. Note that in our simplified framework, $\mA_{m+1}$ is restricted to be chosen from $\setO_F$. We hence employ the \textit{worst-case adaptation} strategy proposed in \cite{muller2018oversampled} and utilized in \cite{muller2018randomoversampled}:~In subframe $m$, the adaptation function finds the permutation 
\begin{align}
\Pi_m \brc{ \dbc{B} } = \set{i_1,\ldots,i_B},
\end{align}
such that $d_{i_1 , m} \geq \ldots \geq d_{i_B , m}$. It then sets the sensing matrix of the next subframe to $\mA_{m+1} \in\setO_F$ whose corresponding~index set is $\setF_{m+1} = \set{i_1,\ldots,i_F}$. The proposed \ac{oas} approach is summarized in Algorithm~\ref{alg1}. %It is worth to indicate that this framework straightforwardly~extends to other prior distributions and models of~structured~sparsity. Due to page limitation, such extensions are skipped here and left for extended versions of the work.

\begin{algorithm}[t]
\caption{Block-Wise OAS via Orthogonal Sensing}
\label{alg1}
\begin{algorithmic}[0]
%\In $\bx$, $K$, $N$, $M > N/K$, $\sigma^2$ and postulated prior $q(x_n)$
\Initiate Set
$d_{b,0} = +\infty$, $\bar{\bw}_b\brc{0} = \boldsymbol{0}_{L\times 1}$ and $\setM_b \brc{0} = \emptyset$.\vspace*{.5mm}
\For{$m \in \dbc{M}$}\\
\begin{enumerate}
\item Determine $\setF_m$ by worst-case adaptation on $\bd_{m-1}$.
\item Update $\setM_b \brc{m} = \setM_b \brc{m-1} \cup \set{m}$ for all $b\in \setF_m$.
\item Select $\mA_m\in\setO_F$ which corresponds to $\setF_m$.
\item Sense the samples for duration $T/M$ via $\mA_m$.
\item Determine sufficient statistic $\bw_m = \mA_m^\trp \by_m$ from  $\by_m$.
\item Update $\bar{\bw}_b\brc{m} = \bar{\bw}_b\brc{m-1} + \bw_{b,m}$ for all $b\in \setF_m$.
\item Update $\hat{\bx}_{b, m}$ and $d_{b,m}$ using \eqref{eq:hat_x_b} and \eqref{eq:d_m} respectively.
\end{enumerate}
\EndFor
%\Out $\hat{\bx}_{b, M}$ for $b \in \dbc{B}$.
\end{algorithmic}
\end{algorithm}

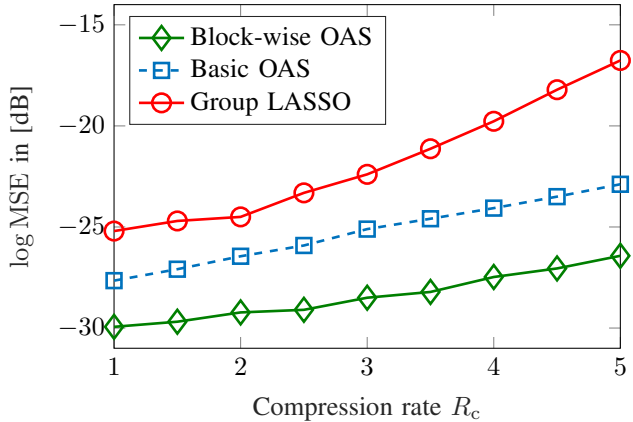
\begin{figure}[t]
\definecolor{mycolor1}{rgb}{0.00000,0.49804,0.00000}%
\definecolor{mycolor2}{rgb}{0.00000,0.44706,0.74118}%
\begin{tikzpicture}

\begin{axis}[%
width=2.65in,
height=1.8in,
at={(1.989in,1.234in)},
scale only axis,
xmin=1,
xmax=5,
xtick={1,2,3,4,5},
xticklabels={{$1$},{$2$},{$3$},{$4$},{$5$}},
xlabel style={font=\color{white!15!black}},
xlabel={Compression rate $R_{\rm c}$},
ymin=-31,
ymax=-14,
ytick={-30,-25,-20,-15},
yticklabels={{$-30$},{$-25$},{$-20$},{$-15$}},
ylabel style={font=\color{white!15!black}},
ylabel={$\log \mathrm{MSE}$ in [dB]},
axis background/.style={fill=white},
legend style={at={(0.03,0.65)}, anchor=south west, legend cell align=left, align=left, draw=white!15!black}
]

\addplot [color=mycolor1, line width=1.0pt, mark size=4.5pt, mark=diamond, mark options={solid, mycolor1}]
  table[row sep=crcr]{%
1	-29.939149213708\\
1.5	-29.6865789330418\\
2	-29.2267295628321\\
2.5	-29.098139025558\\
3	-28.4993877089659\\
3.5	-28.213674101215\\
4	-27.4743834717288\\
4.5	-27.0511183548959\\
5	-26.4257145860491\\
};
\addlegendentry{Block-wise OAS}

\addplot [color=mycolor2, dashed, line width=1.0pt, mark size=2.8pt, mark=square, mark options={solid, mycolor2}]
  table[row sep=crcr]{%
1	-27.6541872434659\\
1.5	-27.0886107246843\\
2	-26.4483081813788\\
2.5	-25.9155185814978\\
3	-25.1044435608676\\
3.5	-24.5921341939462\\
4	-24.0666347980914\\
4.5	-23.4994128556717\\
5	-22.8869256802568\\
};
\addlegendentry{Basic OAS}

\addplot [color=red, line width=1.0pt, mark size=3.5pt, mark=o, mark options={solid, red}]
  table[row sep=crcr]{%
1	-25.2020931401028\\
1.5	-24.7018739439406\\
2	-24.5012144499321\\
2.5	-23.3127540218862\\
3	-22.396151299998\\
3.5	-21.1303398611242\\
4	-19.7708643523274\\
4.5	-18.2095521989687\\
5	-16.761600820679\\
};
\addlegendentry{Group LASSO}

\end{axis}
\end{tikzpicture}%
\caption{MSE vs. compression rate for $B=100$ and $L=4$.}
\label{fig:1}
\end{figure}

\section{Numerical Investigations}
We investigate the proposed framework by conducting some numerical experiments. To this end, we consider the following \textit{time-limited} sensing scenario:
\begin{itemize}
\item $T=1$ and $\sigma^2\brc{t} = 0.01 / t$.
\item The vector of signal samples consists of $B$ blocks of $L$ samples with sparsity factor $\xi = 0.1$.
\item The compression rate is defined as $R_{\rm c} = N/K = BL/K$.
\item The performance is quantified via the \ac{mse}  which is given by the average distortion when $\Delta\brc{\bx; \hat{\bx}} = \norm{\hat{\bx} - \bx}^2/N$.%defined as
%\begin{align}
%\mse = \frac{1}{N} \E \set{ \norm{\hat{\bx} - \bx}^2 }
%\end{align}
%where $\hat{\bx}$ denotes the vector of \textit{final} recovered samples.
\end{itemize}
We study three different signal recovery schemes:
\begin{enumerate}
\item Algorithm~1 with $M=8$ subframes.
\item The \textit{basic} \ac{oas} algorithm with orthogonal measurements which does~not~take the sparsity structure into~account and treats samples as an \ac{iid} sparse Gaussian sequence \cite[Algorithm 1]{muller2018randomoversampled}. Similar to Scheme~1, we set $M=8$.
\item The \textit{benchmark} in which the sensors measure the samples via sensing matrix $\mA$ in a single subframe. The samples are recovered from measurement vector~$\by$ via the \textit{group LASSO} algorithm~which~for~some~$\lambda$ reads \cite{deng2013group}
\begin{align}
\hat{\bx} = \argmin_{\bv\in\setR^N} \left. \norm{\by - \mA\bv}^2 + \lambda \sum_{b=1}^B \norm{\bv_b} \right.
\end{align}
We assume $\mA$ is an \ac{iid} matrix whose entries are zero-mean with variance $1/K$. This is a conventional setting~in classic compressive sensing; see for example \cite{bereyhi2018theoretical,shiraki2016typical}.
\end{enumerate}

Fig.~\ref{fig:1} shows the \ac{mse} against the compression rate $R_{\rm c}$ for all the schemes when $B=100$ and $L=4$. For Scheme 3, the results are given by minimizing the \ac{mse} with respect to $\lambda$ numerically. As the figure shows, the block-wise OAS scheme with $M=8$ subframes outperforms the benchmark for a large range of compression rates. This observation indicates that \textit{even by suboptimal adaptation} the sequential approach of \ac{oas} improves the recovery performance which is intuitive: The proposed algorithm recovers the zero blocks from the low-quality measurements of first few subframes. It then excludes these blocks in next subframes and only measures the non-zero blocks. Due to the longer sensing time, the latter measurements are of higher quality resulting in a good recovery.

\begin{figure}[t]
\definecolor{mycolor1}{rgb}{0.00000,0.49804,0.00000}%
\definecolor{mycolor2}{rgb}{0.00000,0.44706,0.74118}%
\begin{tikzpicture}

\begin{axis}[%
width=2.65in,
height=1.8in,
at={(1.989in,1.234in)},
scale only axis,
xmin=1,
xmax=16,
xtick={1,2,4,8,16},
xticklabels={{$1$},{$2$},{$4$},{$8$},{$16$}},
xlabel style={font=\color{white!15!black}},
xlabel={Block length $L$},
ymin=-29,
ymax=-23,
ytick={-30,-28,-26,-24},
yticklabels={{$-30$},{$-28$},{$-26$},{$-24$}},
ylabel style={font=\color{white!15!black}},
ylabel={$\log \mathrm{MSE}$ in [dB]},
axis background/.style={fill=white},
legend style={at={(.45,0.35)}, anchor=south west, legend cell align=left, align=left, draw=white!15!black}
]
\addplot [color=mycolor1, line width=1.0pt, mark size=4.5pt, mark=diamond, mark options={solid, mycolor1}]
  table[row sep=crcr]{%
1	-24.0134406105169\\
2	-25.7008993599843\\
4	-27.4030609452343\\
8	-28.0518664721055\\
16	-28.2312465981609\\
};
\addlegendentry{Block-wise OAS}

\addplot [color=mycolor2, dashed, line width=1.0pt, mark size=2.8pt, mark=square, mark options={solid, mycolor2}]
  table[row sep=crcr]{%
1	-24.0473299736932\\
2	-24.0671821336786\\
4	-24.0139736130901\\
8	-24.1063191621515\\
16	-23.9862004423665\\
};
\addlegendentry{Basic OAS}

\end{axis}
\end{tikzpicture}%
\caption{MSE vs. block-length for $N=1600$ and $R_{\rm c}=4$.}
\label{fig:2}
\end{figure}
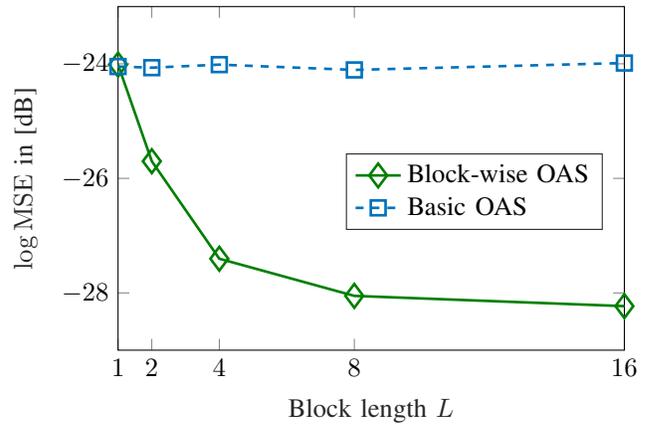

It is further observed in Fig.~\ref{fig:1} that the proposed scheme outperforms the \textit{basic} \ac{oas} algorithm. Such an observation is due to the fact that in \textit{basic} \ac{oas} the sparsity structure does not play any role in the recovery and adaptation. To further illustrate this latter observation, we sketch the \ac{mse} against the block length for Scheme~1 and Scheme~2 in Fig.~\ref{fig:2} when $R_{\rm c} = 4$. To fair comparison, at block length $L$ the number of blocks $B$ is chosen such that $N=1600$. As the figure depicts, the \ac{mse} achieved by  block-wise \ac{oas} reduces as the block-length $L$ increases. This follows the fact that the number of canonical sparse subspaces in the block sparse model reduces with the block length which improves the recovery performance. Such a behavior is however not observed in basic \ac{oas} following the fact that this scheme ignores the sparsity structure.

\section{Conclusions}
A low-complexity \ac{oas} framework has been developed~to sequentially measure signals with structured sparsity. The~proposed scheme exploits the sparsity pattern of the signal~to~improve adaptation and recovery. Our numerical investigations demonstrated that this scheme outperforms the classic non-adaptive compressive sensing framework with the well-known group LASSO recovery algorithm, as well as the basic \ac{oas} framework previously developed for the \ac{iid} sparsity model.

\bibliography{ref}
\bibliographystyle{IEEEtran}
\end{document}